\documentclass[apl,superscriptaddress,reprint,twocolumn,amssymb,aps,floatfix,showkeys]{revtex4-1}
\usepackage{amsmath}%
\usepackage{amsfonts}%
\usepackage{amssymb}%
\usepackage{graphicx}
\usepackage{color}
\usepackage{tabularx}
\usepackage{braket}
\usepackage{wasysym}
\usepackage[tableposition=top]{caption}

\begin{document}

\title{Creating Tuneable Microwave Media from a Two-Dimensional Lattice of Re-entrant Posts}


\author{Maxim Goryachev}
\affiliation{ARC Centre of Excellence for Engineered Quantum Systems, University of Western Australia, 35 Stirling Highway, Crawley WA 6009, Australia}

\author{Michael E. Tobar}
\affiliation{ARC Centre of Excellence for Engineered Quantum Systems, University of Western Australia, 35 Stirling Highway, Crawley WA 6009, Australia}

\date{\today}


\begin{abstract}

The potential capabilities of resonators based on two dimensional arrays of re-entrant posts is demonstrated. {\color{black}Such posts may be regarded as magnetically-coupled lumped element microwave harmonic oscillators, arranged in a 2D lattices structure, which is enclosed in a 3D cavity.} By arranging these elements in certain 2D patterns, {\color{black} we demonstrate how to} achieve certain requirements with respect to field localisation and device spectra. Special attention is paid to symmetries of the lattices, mechanical tuning, design of areas of high localisation of magnetic energy, which in turn creates unique discrete mode spectra. {\color{black} We demonstrate analogies between systems designed on the proposed platform and well known physical phenomena such as polarisation, frustration and Whispering Gallery Modes. The mechanical tunability of the cavity with multiple posts is analysed and its consequences to optomechanical applications is calculated. One particular application to quantum memory is demonstrated with a cavity design consisting of separate  resonators analogous to discrete Fabry-P\'{e}rot resonators.} Finally, we propose a generalised approach to a microwave system design based on the concept of Programmable Cavity Arrays.

\end{abstract}

\maketitle

\section*{Introduction}

Microwave cavities and resonators are important scientific tools used in many physical, chemical, biological and engineering applications to boost system performance. For physical science applications, various cavities are used as sensors\cite{tobar1993parametric,Klein:5aa}, clocks\cite{Locke:2008aa}, spectroscopy\cite{goryachev2013giant,PhysRevB.88.224426,Goryachev:2015ab}, probes of fundamental physics\cite{Stanwix:2005aa,Parker:2013aa} and as building blocks of future quantum systems\cite{Blais:2004aa,Amsuss:2011aa,PhysRevB.90.100404}. These applications utilise the whole range of microwave cavities, either of one-, two- or three-dimension, metallic or dielectric, Fabry-P\'{e}rot type or Whispering Gallery depending on whether on not each of these cavity types meet particular requirements such as: high quality factors, high tunability, energy density in a small volume, separation of magnetic and electric fields, required frequency spectrum, etc. Each of these cavities have advantages and disadvantages with respect to these requirements. In this work we present a new approach to cavity design based on the recently proposed multi-post re-entrant cavity\cite{Goryachev:2015aa,patent2014} that is flexible enough to meet all these criteria simultaneously together with extra new features. {\color{black} Multi-post re-entrant cavities are generalisation of well known one-post reentrant cavities\cite{reen0,reen1,reen2} characterised by high mechanical tunability\cite{Carvalho:2014aa} and potentially high Quality factors for superconducting realisation\cite{Bassan2008}. They are widely used as ultra-sensitive transducers for example in Gravitational Wave antennas\cite{Blair:1995zr,Aguiar:2006aa}, microwave enhanced chemistry\cite{Kalhori}, dielectric measurements\cite{nist1}, gas and liquid characterisation\cite{May:2002aa}, etc.}

In the following discussion, we deal with closed metallic cavities having metallic rods from the top surface to the bottom. These rods are referred as posts, and a possible small distance between the rod and the bottom surface is called the gap. Each post equipped with a finite gap can be viewed as a lumped Harmonic Oscillator with the post body being an equivalent inductance and the gap being the equivalent capacitance. Posts located in the vicinity of each other also have some mutual inductance that introduces coupling between the corresponding HO. {\color{black}The posts may be arranged in different two dimensional patterns or lattices. For this reason, the lattice is referred to as  2D though the original one-post re-entrant resonator is a cavity that belongs to the class of 3D cavities. Indeed, when a multi-post generalization is constructed, it is done in only two dimensions since the dimension along posts is vacuous and is essentially different from the other two. The same logic is used for the designation of, for example, microstrip circuits as 2D systems: although they are embedded into 3D structure, one of the dimensions is different in nature and the circuit design is made in two dimensional space. For this reason, all cavities in this work are shown only in two dimensions depicting the working plane.
}

The multipost re-entrant cavity has much in common with both the 3D cavities and lumped component resonators. Having the advantage of high quality factors specific to 3D closed cavities, it has the advantages of discrete lumped circuits such as high concentration of the electric energy and high tunability. Moreover, it has been demonstrated that these cavities may be used to achieve high concentrations of the magnetic field\cite{Goryachev:2014aa,Creedon:2015aa} that is useful for many spectroscopy applications.
The uniqueness of the cavity is due to the discreteness of its elementary parts (metallic posts). Indeed, if the cavity is made of a good conductor, each post may be assigned with the specific direction of the current at each moment of time. Combinations of these directions distinguishes different modes of the cavity. As a result, the cavity design is reduced to an arrangement of posts on a two dimensional lattice and investigation of the cavity properties with respect to different 2D arrangements. {\color{black} The purpose of this work is to demonstrate various aspects of the 2D structures of posts and establish intuition for such cavity design as well as draw analogies with existing physical systems. For this reason, the work describes mode polarisation and frustration phenomena as well formation of discrete Whispering Gallery and Fabry-P'{e}rot Modes. Additionally, we describe one of the main features of re-entrant cavities namely mechanical tuning applied to multi-post structures and its consequences for a typical optomechanical problem. Then, we give an example of a multipost structure that may be used as a quantum memory, and, finally, we introduce an idea of programmable cavity arrays as a generalisation of the cavity design approach.}

\section{Current Frustration}

\begin{figure}[t!]
	\centering
			\includegraphics[width=0.4\textwidth]{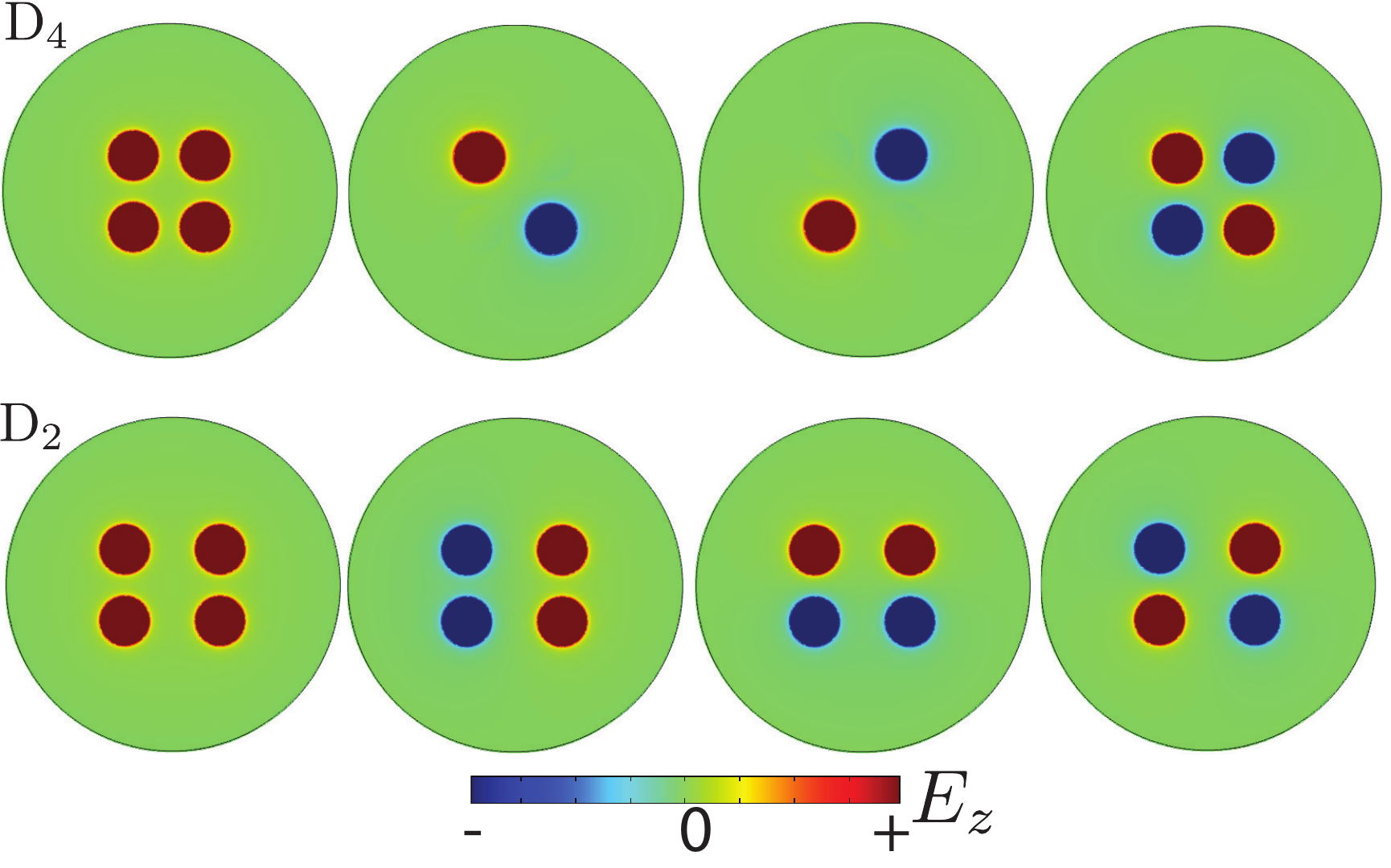}
	\caption{Fields patterns for cavities with $N=4$ and different symmetries. The colour shows the direction of the electric field in the gap along the cavity height.}
	\label{D4D2}
\end{figure}

A multipost cavity can be regarded as a system of coupled Harmonic Oscillators (HO). Indeed, each post represent an electrical HO with the gap playing the role of a capacitor and the post itself being an inductor. The cavity exhibits a number of resonance frequencies corresponding to the number of posts $N$. Each resonance is uniquely identified with a pattern of current directions (at the same instance of time) giving a unique magnetic field distribution\cite{Goryachev:2015aa}. Fo example, a two post cavity\cite{Goryachev:2014aa,Creedon:2015aa} exhibits two re-entrant modes: a dark mode ($\uparrow\uparrow$ mode) with the both posts having the same direction and a bright mode ($\uparrow\downarrow$ mode) with the opposite direction of currents at any instance of time. The same logic and description may be applied to more complicated cases. Although, more generally the result depends on the cavity symmetries. In the case of a cavity with $N=4$ arranged in a rectangle (D$_2$ symmetry) the modes are organised in the following order starting from the lower to higher frequency:
$\uparrow\uparrow\uparrow\uparrow$ - $\uparrow\uparrow\downarrow\downarrow$ - $\uparrow\downarrow\downarrow\uparrow$ - $\uparrow\downarrow\uparrow\downarrow$. Since all posts are identical, the second and the third modes are identical in frequency since they are related via a symmetry operation ($\pi/2$ rotation). The comparison between two $N=4$ objects with different symmetries is shown in Fig.~\ref{D4D2}. It should be mentioned here that objects with the same symmetry type but different geometry can exhibit different field patterns. 

In some cases a situation arises when for symmetry reasons it is not possible to decide on the pattern of the current for some modes. For instance in the case of a three post cavity with the D$_3$ symmetry, the cavity's two higher frequency modes cannot be identified with any possible current pattern. Indeed, if two posts have $\uparrow\downarrow$ current direction, the third node cannot be associated with any direction due to symmetry reasons and one post cannot sustain two current directions simultaneously as well. Regardless the choice, the system has an asymmetric solution. The same holds true if one chooses $\uparrow\downarrow$ as a pattern for first two posts. Such modes we label as 'frustrated' modes or modes exhibiting current frustration in analogy with spin frustration\cite{Pekalski:1980aa} well known in condensed matter physics. Table~\ref{T1} summarises these properties for some low order cases. 

It has to be underlined that in practice, the current frustration cannot be achieved. The reason is due to unavoidable symmetry breaking arising from any cavity imperfections. Moreover, even for Finite Element Modelling (FEM), the symmetry is broken by small asymmetries of the mesh. Although, in this case, this property can be easily controlled by the mesh size.


\begin{table}
\caption{Classification of some cavities and corresponding modes. '0' is the canceled (for symmetry reasons) current. '?' denotes frustrated current. The last column demonstrates wether the system has degenerate modes or not.}
\begin{tabular}{r|c|c|c|c}
\hline
$N$&Symmetry&Modes&Frustrated?&Degenerate?\\\hline\hline
2& D$_1$ & $\uparrow\uparrow$-$\uparrow\downarrow$&No&No \\\hline
3& D$_3$ & $\uparrow\uparrow\uparrow$-$\uparrow\downarrow?$-$\uparrow?\!\uparrow$&Yes& \\\hline
3& D$_1$ & $\uparrow\uparrow\uparrow$-$\uparrow\!0\!\downarrow$-$\uparrow\downarrow\uparrow$&No&No \\\hline
4& D$_4$ & $\uparrow\uparrow\uparrow\uparrow$-$\uparrow\!0\!\downarrow\!0$-$0\!\downarrow\!0\!\uparrow$-$\uparrow\downarrow\uparrow\downarrow$&No&Yes \\\hline
4& D$_2$ (rect.) & $\uparrow\uparrow\uparrow\uparrow$-$\uparrow\uparrow\downarrow\downarrow$-$\uparrow\downarrow\downarrow\uparrow$-$\uparrow\downarrow\uparrow\downarrow$&No&No \\\hline
5& D$_5$ & $\uparrow\uparrow\uparrow\uparrow$-$\uparrow\uparrow\downarrow\downarrow?$-$\uparrow\uparrow\downarrow\downarrow?$&Twice& \\
&  & $\uparrow\downarrow\uparrow\downarrow?$-$\uparrow\downarrow\uparrow\downarrow?$&& \\\hline
\end{tabular}
\label{T1}
\end{table}   

\section{Discrete WGMs}

Many of the resonant system may be subdivided into two groups: Fabry-P{\'e}rot (FP) and Whispering Gallery Mode (WGM) resonators. The first type have their largest wave-vector situated on a segment of a line. Experimentally, the segment is formed by two boundary conditions limiting the wave propagation medium. 
The second cavity type is characterised by a wave-vector defined on a circle with the rotational symmetry in the ideal case. The wave is formed due to full internal reflection form the media boundary. It has been demonstrated\cite{Goryachev:2015aa} that arrays of posts can make media for Fabry-P{\'e}rot type resonances with both optical and acoustic branches of a dispersion relation. In the same manner, for $N>2$, D$_N$ cavities may be considered as discrete WGM resonators. 

For a resonator defined on a 1D structure (circle or segment) of posts, one can introduce a wave length $\lambda$ as a number of posts covered by one complete variation of currents (see \cite{Goryachev:2015aa} for illustrations in the case of a segment). Though the system is defined on a discrete set $\lambda$ can be fractional. For instance if a resonator is constructed of $N$ posts, its mode numbers $n$ should run from 1 to $N/2$. Obviously, not all these numbers are integers. Fractional wavelengths correspond to frustrated modes discussed above. In the case of the broken symmetry, the fractional wavelength modes appear as modes with variable wavelength.  

\begin{figure}[t!]
	\centering
			\includegraphics[width=0.4\textwidth]{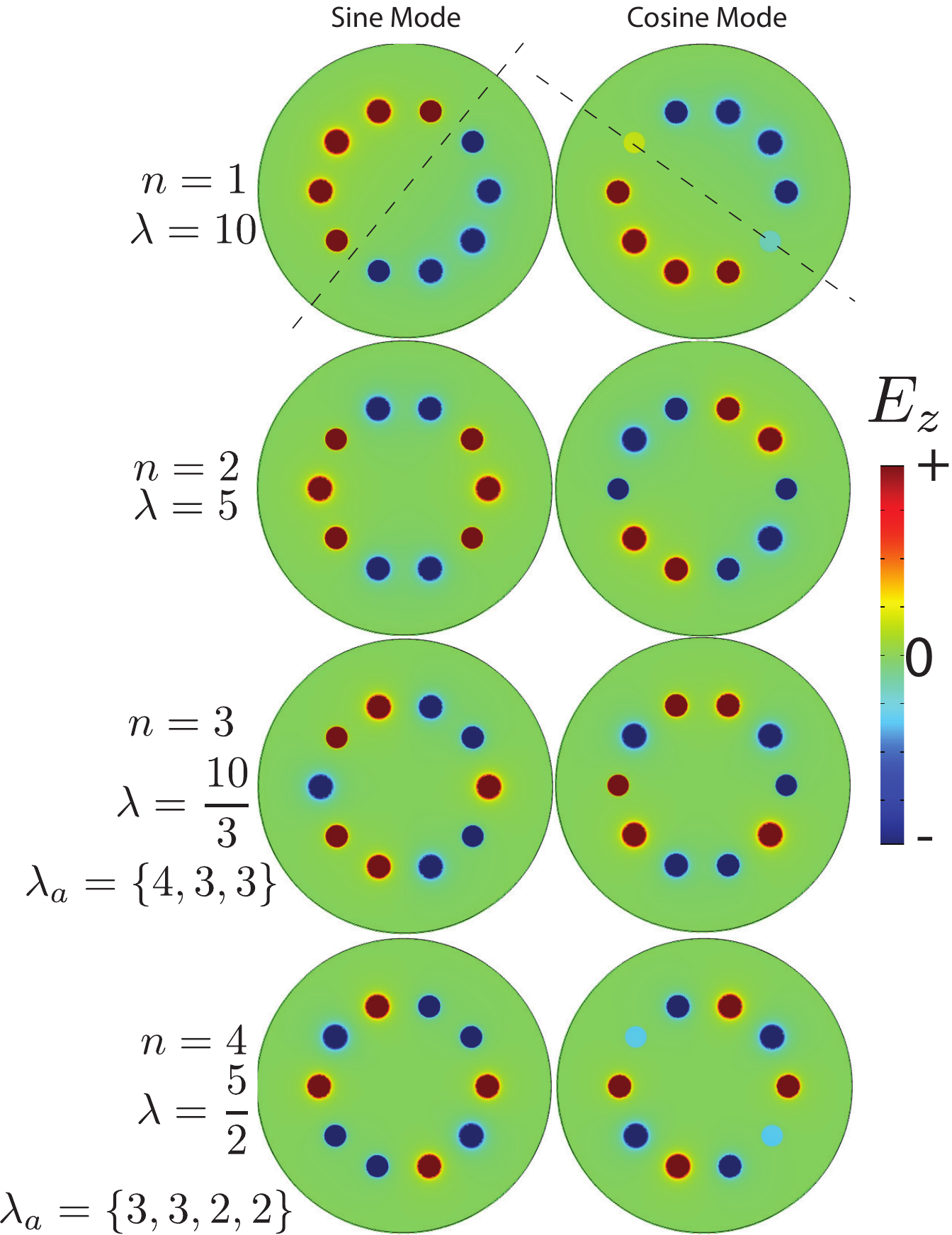}
	\caption{FEM simulation of doublet modes of a D$_{10}$ cavity. The colour shows the direction of the electric field in the gap along the cavity height. Modes are characterised by the mode number $n$ and wave length $\lambda$. $\lambda_a$ are actual wavelengths.}
	\label{doublet}
\end{figure}

Similarly to traditional WGM devices, for discrete WGMs, one can introduce a wave number along the circle containing the corresponding regular polygon. Moreover, because of the rotational symmetry of the system, the multiple post cavities demonstrate so-called doublets\cite{PhysRevA.89.013810}. WGM doublets are often observed in high $Q$-factor cylindrical resonators such as sapphire single crystals\cite{Bourgeois:2005aa} {\color{black}or nonuniform circular dielectric resonators\cite{Filipov:1995aa}}. Figure~\ref{doublet} illustrates this property of a FEM approximation to the D$_{10}$ cavity resonator. This type of the WGM cavity shows eight doublet modes where four modes have variable wavelengths. Two more modes $\uparrow\uparrow\uparrow\uparrow\uparrow\uparrow\uparrow\uparrow\uparrow\uparrow$ ($N=0$) and $\uparrow\downarrow\uparrow\downarrow\uparrow\downarrow\uparrow\downarrow\uparrow\downarrow$ ($N=10$) are identical under all cavity symmetry operations. 

Another feature of the discrete WGM cavity is associated with its symmetry. Since the cavity does not have the continuous rotational symmetry, the sine and cosine components of the doublet are not equivalent by default. This property is clearly seen for the first doublet in Fig.~\ref{doublet}.


Generalising these results, the re-entrant cavity modes for an D$_N$ system perform an arithmetic expansion of the integer $N$. The expansion is made by the cavity wavelengths because of the fact that $N$ is constructed by $n$ wavelengths (mode order) of length $\lambda$. When $N$ is not devisable by $n$, $\lambda$ is variable and made by a set of different numbers $\lambda_a$. Generally, the sum of all wavelengths $\lambda_a$ is $N$. This principle is demonstrated for D$_{10}$ in Fig.~\ref{doublet}, where the cavity reveals 9 of such expansions (the first mode with all currents in the same direction may be understood as a 0-order mode with the infinite wavelength). Note that these expansions are not all expansions of $N$ into a sum of integers, but only those with terms smaller than $N/2$. In other words, the cavity performs division of an integer $N$ by $n\in\{0..N/2\}$.

\section{Mode Polarisation}

\begin{figure}[t!]
	\centering
			\includegraphics[width=0.48\textwidth]{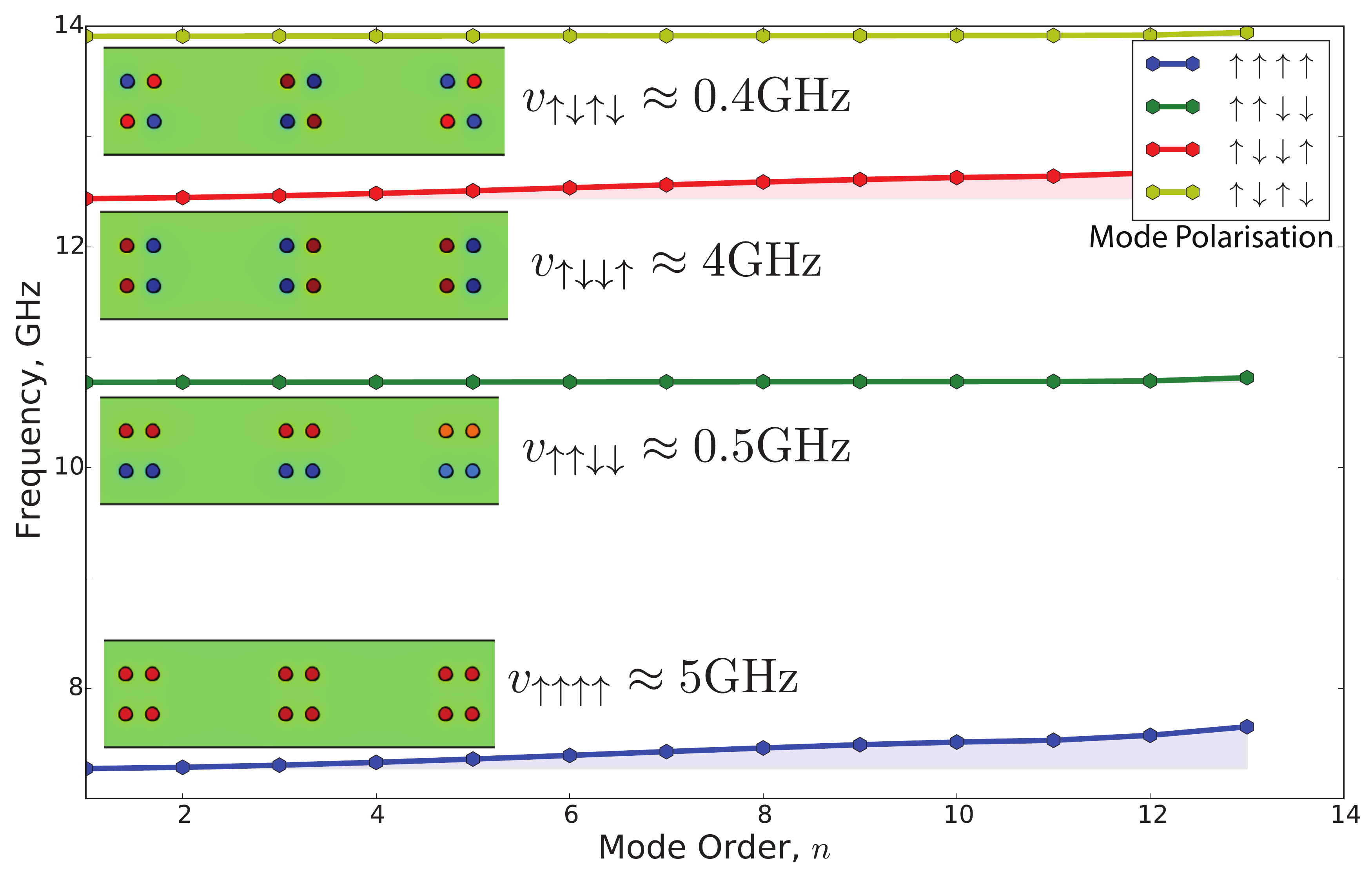}
	\caption{Dispersion curves for one dimensional array of 4-post D$_2$ cavities. The modes are classified into four curves with different polarisations. The difference between acoustic and optical branches is neglected. $v$ denotes the estimated speed of propagation.}
	\label{polar}
\end{figure}

It has been already demonstrated\cite{Goryachev:2015aa} that one dimensional array of post can simulate quasi particles on finite lattice. The same structure can be used to demonstrate the concept of bandgaps and various impurities. To generalise this concept, it is possible to use some primitives of posts arrangements. For example, one can take the D$_2$ arrangement of four posts and analyse the dispersion relationship of one dimensional arrangement of these structures. The dispersion curves for such system are shown in Fig.~\ref{polar}. All modes are classified into four groups having different polarisations. These polarisations correspond to different modes of the primitive cavity (see Table~\ref{T1}). So, the analysed media of D$_2$ primitives exhibit modes with four types of polarisation. Each type is characterised by a different energy in the long wavelength limit and different speed of propagation $v$. For the discrete resonator, this quantity is measured in units of frequency if distances are measured in numbers of lattice cites   

The demonstrated principle is based on the fact that the energy of interaction between two posts depends on the distances between them. So, closer situated posts are characterised by higher "bonding" energies. As a result, any change in the primitive cell results in much larger change in frequency than any change along the larger 1D array structure. The same energy level scaling happens in Nature, where laws of physics can be studied at molecular, atomic, nuclear, etc levels. In this sense each D$_2$ structure in the example may be associated with an atom. Further, such atoms are arranged in an 1D lattice. 

\section{Mechanical Tuning of a Multiple Post Cavity}

\begin{figure}[t!]
	\centering
			\includegraphics[width=0.48\textwidth]{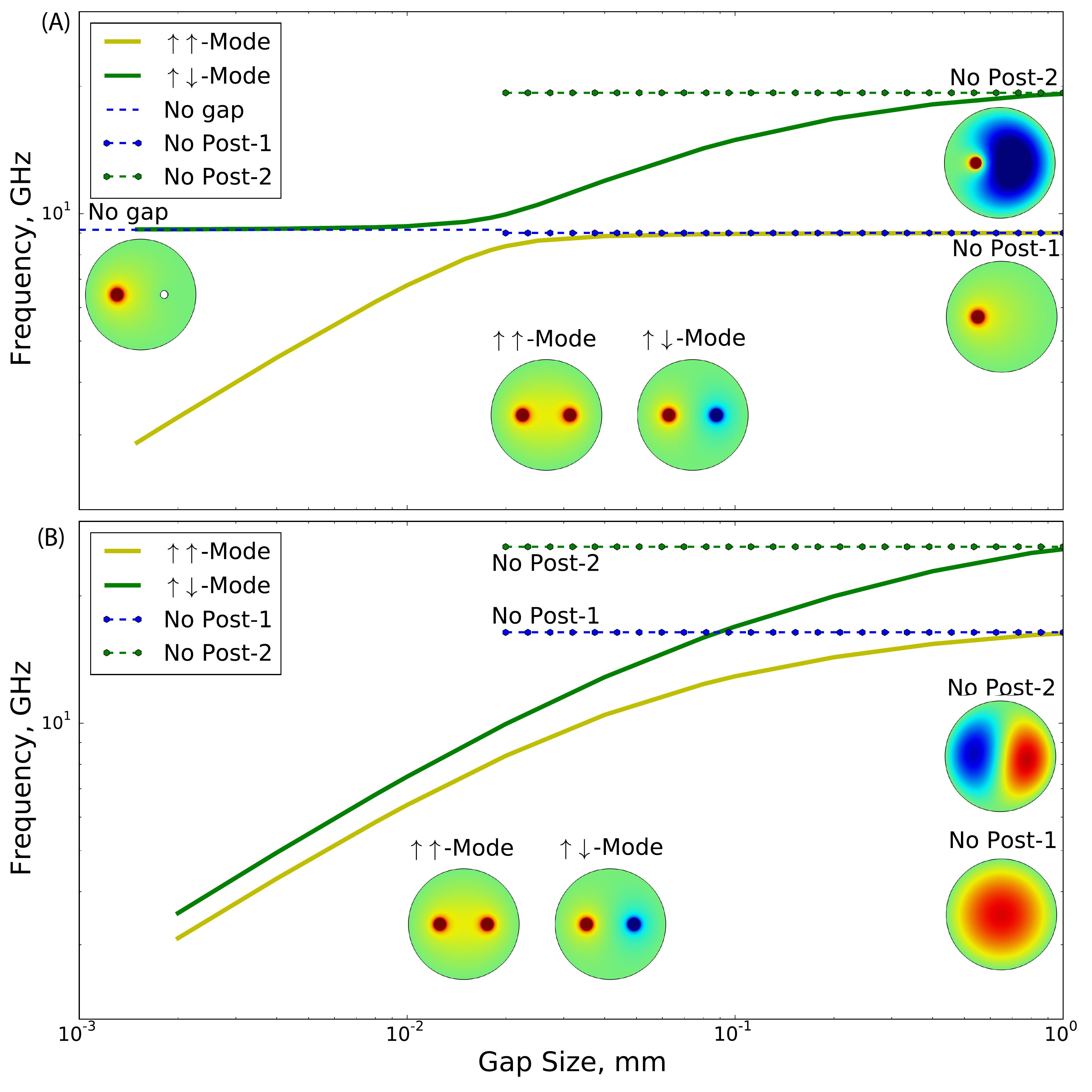}
	\caption{Sensitivity of the two-post cavity to the gap size: (A) only one gap is tuned, (B) both gaps are tuned. Additionally the figure above shows two limiting cases (dashed asymptotes): 1) no gap case (the case where the controlled gap is decreased to 0 making the post complete), 2) no post case (the controlled gap is of the size of the post, so that it does not exist). The insets show distribution of the normal component of the electric field in the resonator plain.}
	\label{modes1}
\end{figure}

\begin{figure}[t!]
	\centering
			\includegraphics[width=0.48\textwidth]{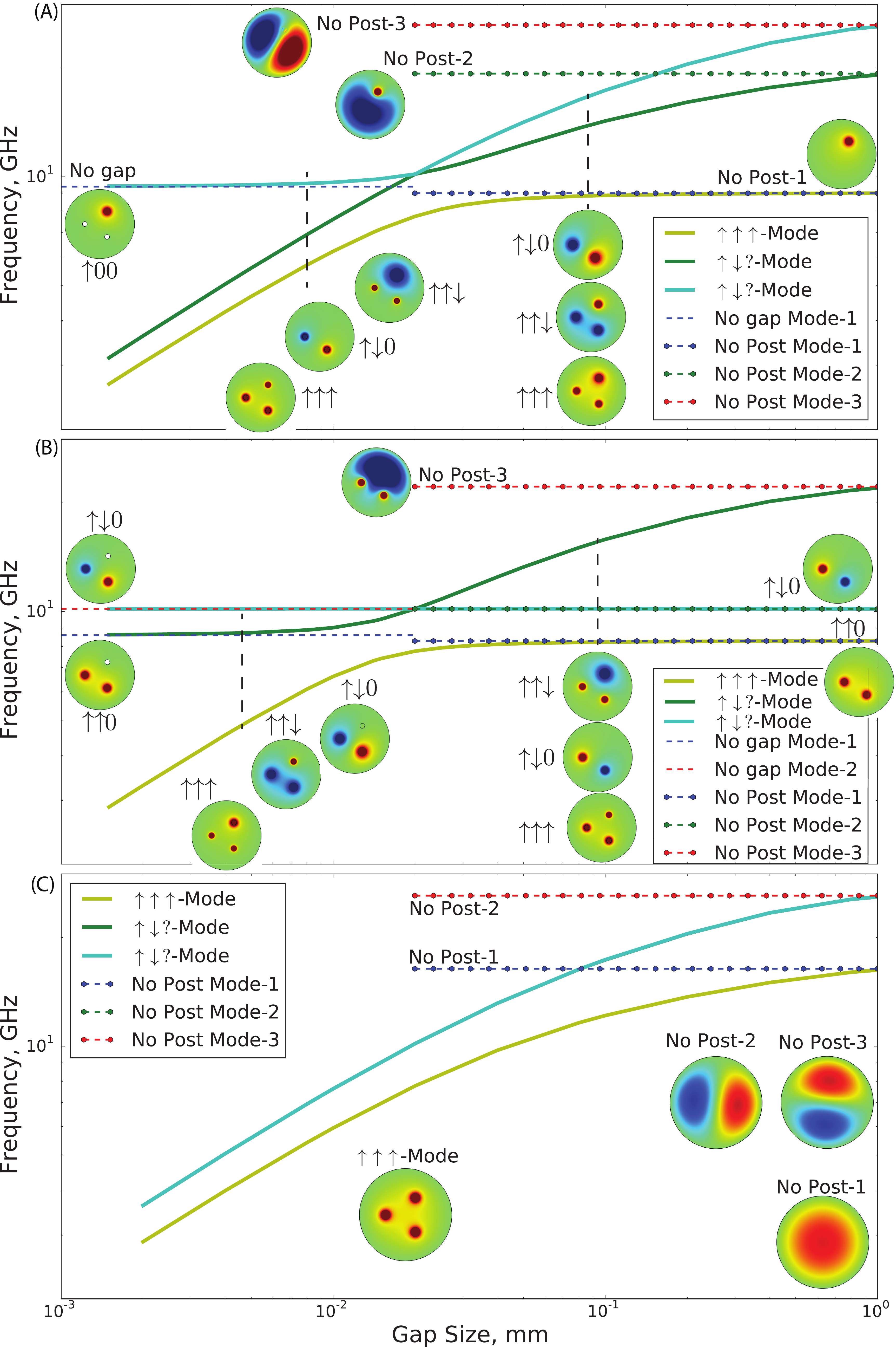}
	\caption{Sensitivity of the three-post cavity to the gap size: (A) one gap is tuned, (B) two gaps tuning, (C) all gaps tuning. Limiting cases are shown with dashed asymptotes. The intersections between curves correspond to the frustration case, where the system has D$_2$ symmetry. For the (C) case, two modes are always degenerate. The insets show distribution of the normal component of the electric field in the resonator plain.}
	\label{modes2}
\end{figure}

One of the main advantages of the re-entrant cavity is its high mechanical tunability with the post gap. This type of tuning has an advantage of speed over the more traditional magnetic field tuning. This property has been already demonstrated for a single post system\cite{reen1}. For a multiple post cavity, this property has not been discussed yet. So, in order to demonstrate this property, a series of FEM simulations have been performed for a 2-post D$_1$ (Fig.~\ref{modes1}) and 3-post D$_3$ cavities (Fig.~\ref{modes2}). 

Fig.~\ref{modes1} shows the dispersion curve for re-entrant modes for $N=2$ case, when only one gap is tuned (A) and both gaps are changed (B). The figure also demonstrates modes in two limiting cases (dashed asymptotes): no post case (the gap reduces the post completely) and no gap case (the gap height is zero). In the case of one-gap tuning, the system exhibits an anti-crossing between two modes. The point of minimal distance between the curves is the symmetry point when both gaps are equal. When one gap decreases, one of the mode frequency converges to zero giving only one re-entrant mode in the no gap limit. For the case of two gaps tuning, the system is always symmetric and two curves do not demonstrate an anti-crossing. Moreover, for small gaps the two curves are almost parallel demonstrating the same sensitivity to the gap size. 

Simulation results for the D$_3$ cavity case demonstrated in Fig.~\ref{modes2} are organised in three groups: (A) one, (2) two and (C) three gap tuning. In one and two gap tuning cases, the system exhibits an anti-crossing between the curves. In both cases, the higher frequency modes cross at the point of the minimal splitting with the lowest frequency mode. This point corresponds to the symmetrical case. In practice, the symmetry is always broken due to small cavity imperfections. These imperfections couple the two modes giving rise to another anti-crossing. In the no gap case, the system is simply reduces to a case with lower $N$ ($N=1$ for (A) and $N=2$ for (B)). By passing the symmetry point, the current structures of two 'degenerate' modes are swapped.
In the case, where all three gaps are tuned, two higher frequency modes stay always degenerate (D$_3$ is always preserved). All modes have the same tuning coefficient at relatively low gap size.

In all the cases described above, a number of modes that are actually tuned equals the number of controlled posts. Generally the system may be represented as a collection of coupled and uncoupled (degenerate mode) HOs. The gap tuning changes the current structure, and thus the cavity field patterns.  
These observations can be generalised onto a cavity with any number of posts. 

The described situation when one cavity mode has a large mechanical tuning coefficient and another is completely insensitive to this mechanical motion is important for practical sensing applications. Indeed, since both modes share the common environment except for the external mechanical force, one can design a dual mode system with all unwanted external effects compensated.

On the other hand, as it is demonstrated in Fig.~\ref{modes1} and \ref{modes2}, when all gaps are synchronously controlled all modes show the same sensitivity. So, in the case when optomechanical coupling exceeds both cavity and mechanical resonator linewidths as well as the cavity free spectral range, one may obtain 'optomechanical superstrong coupling', the optomechanical version of the atom-cavity superstrong coupling regime described by optical cavity Quantum Electrodynamics \cite{Meiser:2006aa}.

\section{Coupling to Higher Order Mechanical Modes}

The result of the previous section may be understood as sensitivity of microwave coupling to motion of some parts of the cavity wall (a conducting membrane) under the posts or its part. The corresponding mechanical motion represent the first order mode where all parts of the cavity wall move synchronously. For the higher order modes, parts of the membrane could move in opposite directions. In this section, we consider the case when two posts are placed above antinodes of the second order membrane mode. Thus, displacements exhibiting by the gaps are equal and opposite at any time. 

\begin{figure}[t!]
	\centering
			\includegraphics[width=0.48\textwidth]{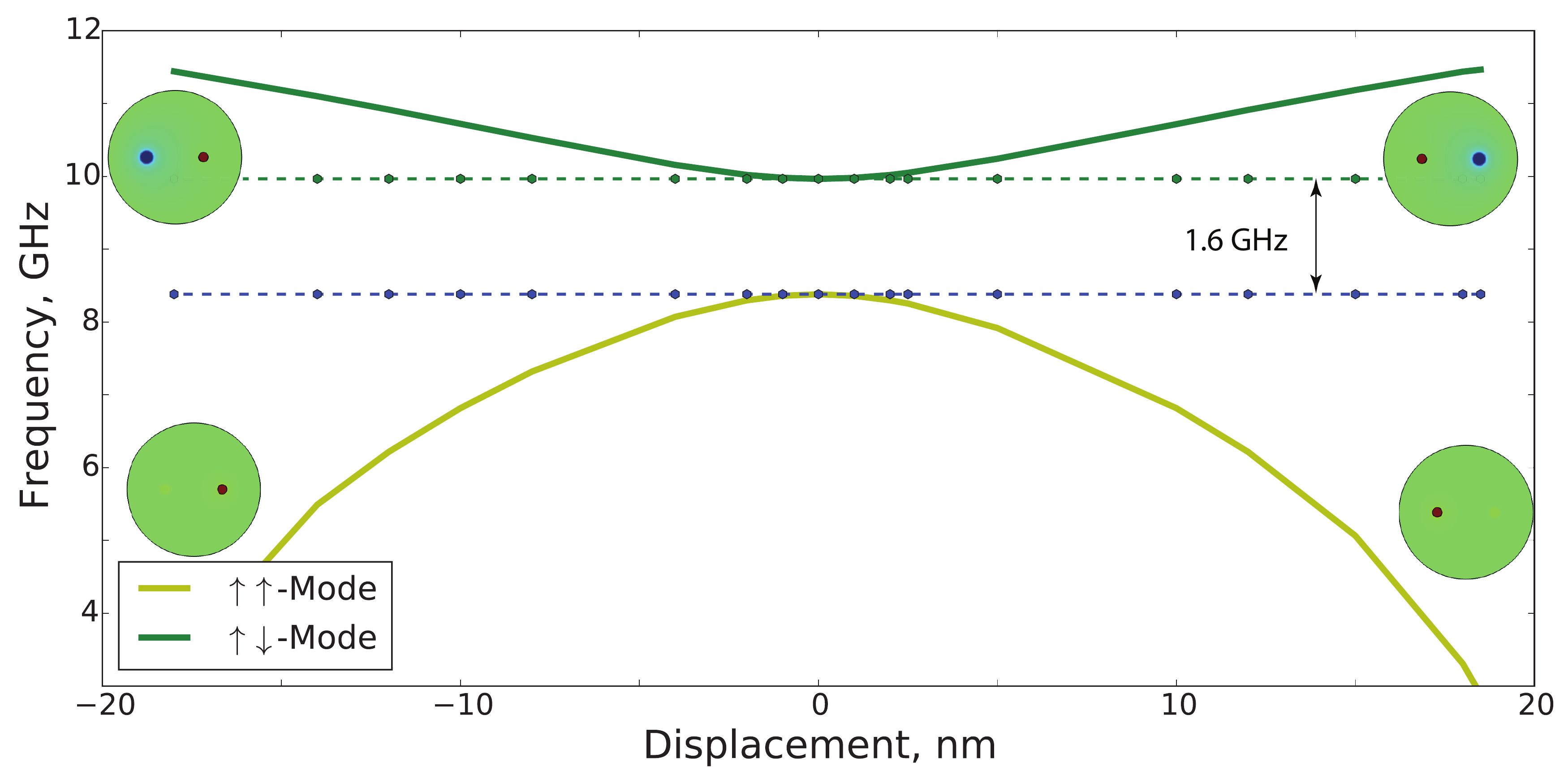}
	\caption{Sensitivity of the two post cavity to the second order displacement.}
	\label{secorder}
\end{figure}

The sensitivity curves for the double post system with the second order displacement is shown in Fig.~\ref{secorder}. The result may be interpreted as mechanically controllable between the dominant post leading to an avoided level crossing. The coupling strength is the function of the mode splitting in the symmetric case. 

\section{Doublet Mode Optomechanics}

It has been demonstrated that D$_{N}$ cavities demonstrate the phenomenon of WGM doublets similar to that of cylindrical dielectric single crystal cavities\cite{Bourgeois:2005aa}. This phenomenon is very common for high quality factor cavities. It is observed when the symmetry breaking coupling between two degenerate solutions exceeds the mode bandwidth. Though the double mode splitting in single crystal cavities may be controlled via an artificial defect\cite{Mazzei:2007aa} or via a paramagnetic spin ensemble with an external magnetic field\cite{PhysRevA.89.013810}, the resulting system flexibility and dynamic properties (slow). On the other hand, the multi-post re-entrant cavity proposes a better solution due to its high mechanical tunability described in the previous sections. Moreover, the setup may serve as a base for multimode microwave optomechanics with large electro-mechanical coupling coefficients. In this case, the cavity wall under the post gap serves as a mechanical membrane resonator parametrically coupled to the cavity modes. The parametrical coupling comes from mechanical motion modulating the post gaps and as a result the equivalent capacitances. Ignoring higher order terms in the electromechanical coupling, the system Hamiltonian could be written as follows:
  \begin{equation}
	\label{J120PJ}
	 \left. \begin{array}{ll}
\displaystyle H = \left(g \left({{b}^\dagger} + {b}\right) + \omega_{c}\right) \left({{a_R}^\dagger} {a_R} + {{a_L}^\dagger} {a_L}\right)+ \omega_{m} {{b}^\dagger} {b}+\\
\displaystyle G \left({{a_R}^\dagger} {a_L} + {a_R} {{a_L}^\dagger}\right) + \left(A e^{i \omega_{d} t} + A e^{- i \omega_{d} t}\right) \left({{a_R}^\dagger} + {a_R}\right) 
\end{array} \right. 
\end{equation}
where $a_R^\dagger$ and $a_L^\dagger$ ($a_R$ and $a_L$) are creation (annihilation) operators for both modes of the doublet of the frequency $\omega_c$ and the interlude coupling $G$, $b^\dagger$ ($b$) is the creation (annihilation) operators of the membrane, $A$ and $\omega_d$ are the amplitude and circular frequency of the driving field. Applying the usual Rotating Wave Approximation and removing coherent part of the state, the linearised Hamiltonian in the limit of strong driving is
  \begin{equation}
	\label{J121PJ}
	 \left. \begin{array}{ll}
\displaystyle H =  \Delta \left({{a_R}^\dagger} {a_R} + {{a_L}^\dagger} {a_L}\right)+ \omega_{m} {{b}^\dagger} {b}+\\
\displaystyle +G \left({{a_R}^\dagger} {a_L} + {a_R} {{a_L}^\dagger}\right) + \alpha^2 g \frac{\Delta^2+G^2}{\Delta^2}(b+b^\dagger)+\\
\displaystyle + \alpha g \Big[\frac{G}{\Delta}(a_L +a_L^\dagger)(b+b^\dagger)-(a_R +a_R^\dagger)(b+b^\dagger)\Big]
\end{array} \right. 
\end{equation}
where $\alpha =  \frac{A\Delta}{\Delta^2-G^2}$, $\Delta = \omega_c-\omega_d$ is the detuning frequency. Note that by putting $G=0$, i.e. in the symmetric case, Hamiltonian (\ref{J121PJ}) reduces to the usual one mode linearised Hamiltonian. 

\begin{figure}[t!]
	\centering
			\includegraphics[width=0.48\textwidth]{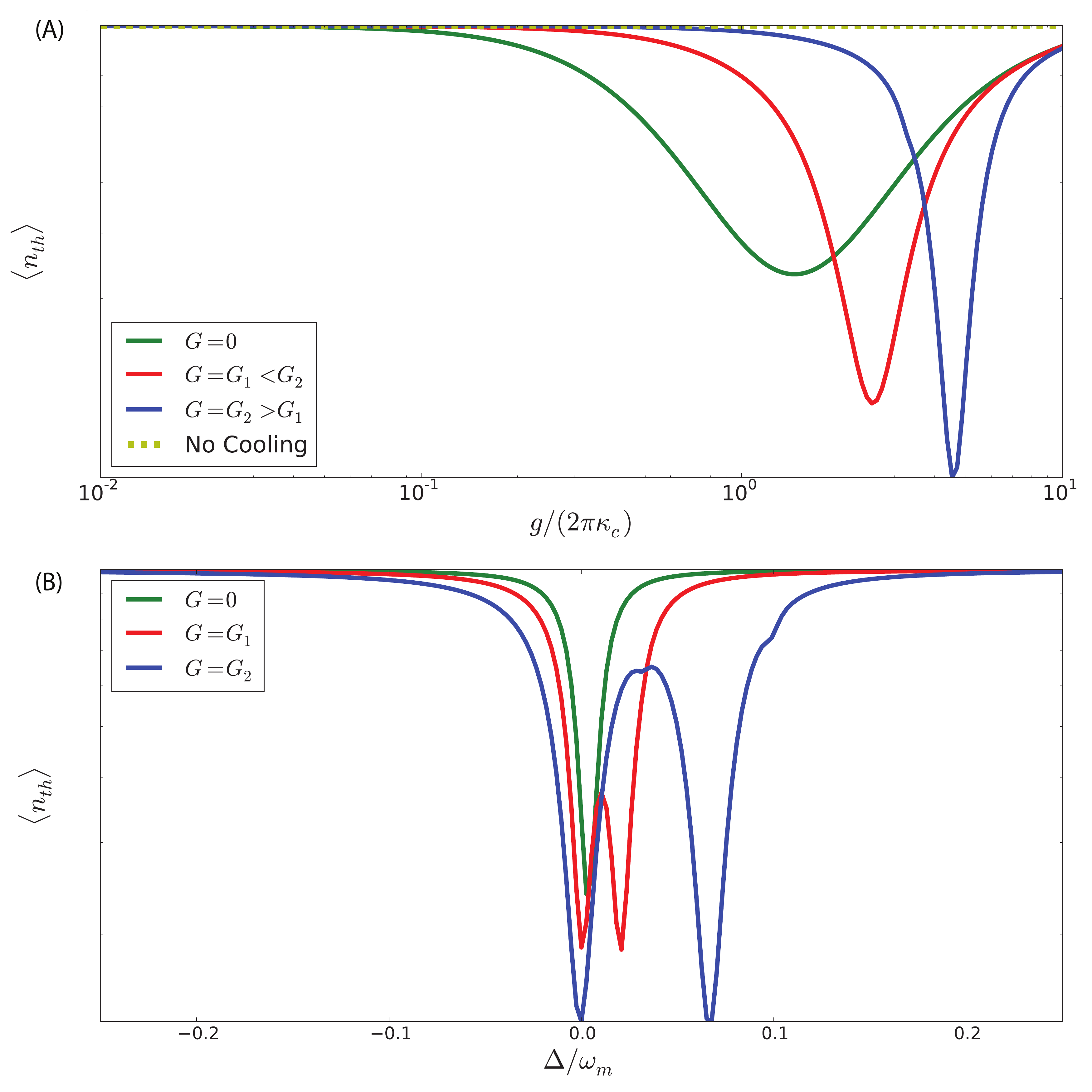}
	\caption{Occupation number of the mechanical membrane mode coupled to a doublet mode of a multi-post cavity as a function of electro-mechanical coupling $g$ (A) and detaining frequency $\Delta$ (B) for various photon-photon couplings $G$.}
	\label{cooling}
\end{figure}

The mechanical mode cooling properties the doublet mode system in the red sideband detuning regime are shown in Fig.~\ref{cooling}. The result is obtained via numerical simulation\cite{Johansson20131234} of the system with certain realisable parameters. Plot~\ref{cooling}, (A) shows cooling improvement over a certain electromechanical-coupling range, at the same time, the doublet mode coupling leads to the splitting of the occupation number minimum of the mechanical mode as shown in Fig.~\ref{cooling}, (B).

\section{Array of Resonators}

\begin{figure*}[t!]
	\centering
			\includegraphics[width=0.75\textwidth]{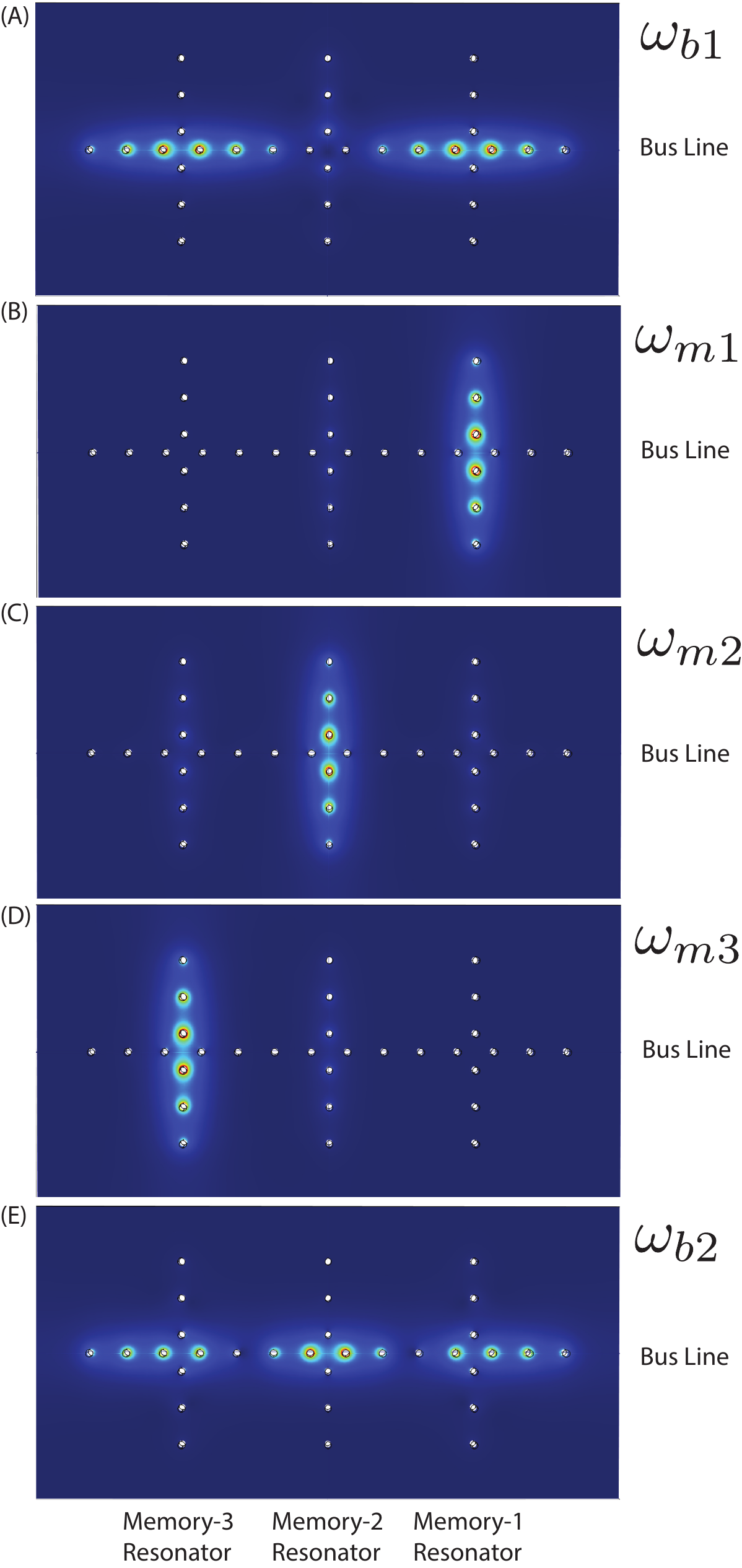}
	\caption{Density plots (absolute value of the magnetic field strength $|H|$) of a bus ((A) and (E)) and memory modes ((B)-(D)) of the register.}
	\label{register}
\end{figure*}

As it is discussed in the previous sections, by combining posts, it is possible to create a discrete approximation to any kind of 1D or 2D cavities. Moreover, it is possible to organise these second level subsystem into arrays creating the next level of organisation. For instance, it is possible to design a system with several "discrete" Fabry-P\'erot cavities  with close enough resonances, such that it is possible to control their frequency individually. In principle, this feature may be used for quantum state transfer and memory. 
As an illustrative example, we consider a cavity that consists of four 1D re-entrant sub-lattices: one bus resonator (bus line) and three memory resonators. The system is designed in such a way that the bus line supports supports two resonances (in this case - second and third harmonic of a line, $\omega_{b1}$ and  $\omega_{b2}$) and memory modes (fundamental resonances of three independent lines $\omega_{m1}$-$\omega_{m3}$). For this particular design $\omega_{b1}<\omega_{m1}\sim\omega_{b2}\sim\omega_{m3}<\omega_{b2}$. No other system mode is located in between. The difference between resonance frequencies of the memory modes is primarily due to imperfections of the numerical model (mesh) breaking the system symmetries. The distribution of the magnetic field for these modes is shown in Fig.~\ref{register}. 


The results of FEM simulation (Fig.~\ref{register}) show that the five modes described above may be regarded as normal modes of the stand alone sub-systems, e.g. the bus or memory resonator-1. This come from the fact that none of the foreign posts are illuminated while one of these subsystem is at the resonance giving all the magnetic field is concentrated in the corresponding regions of this subsystem. Thus, small regions of the cavity space may be addressed individually at the corresponding resonance frequencies, so it is possible to put spin ensembles or qubits at these positions for individually addressing.

Note that none of the sub-resonators of the system share the same posts. So, in principle, it is possible to control the resonance frequencies of the memory and bus modes individually through mechanical manipulation of the corresponding  post gaps. So, it makes it possible to exchange excitations between the bus and memory modes by tuning their frequencies through each other.  
 
\section{Mechanically Programmable Cavity Arrays}

\begin{figure*}[t!]
	\centering
			\includegraphics[width=0.8\textwidth]{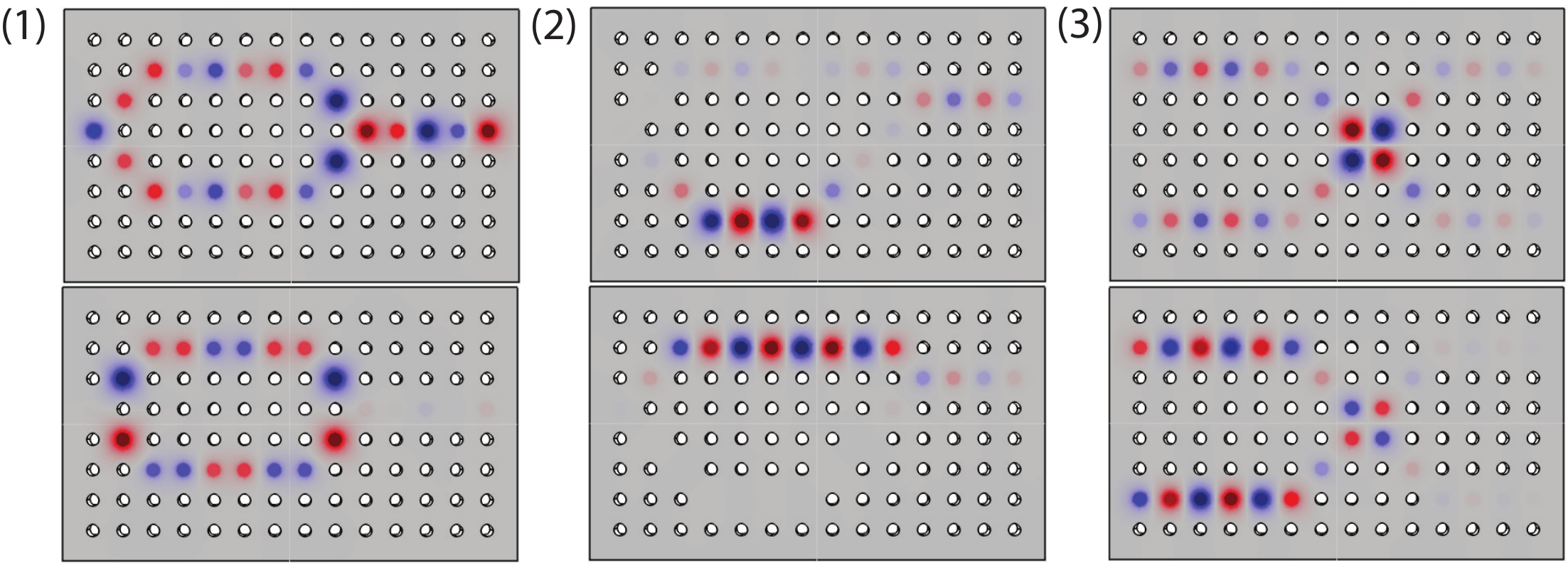}
	\caption{Transmission patterns for three different arrangements of the programmable cavity array. The density plots show the electric field in the cavity plane (under the posts) along the post direction.}
	\label{MPCA}
\end{figure*}

Fig.~\ref{modes1}, \ref{modes2}, \ref{secorder} demonstrate a remarkable feature of the re-entrant post cavity: a system resonance is associated only with a post with a gap. When a post gap disappears, the resonance frequency approaches zero. This opportunity may be utilised to control the number of resonators and their geometry. For this purpose, one creates a regular grid of posts where each post may or may not have a gap. Then, by assigning gaps only to certain posts, one can create chains and other structures of coupled oscillators on a 2D array. So, due to the re-entrant cavity feature, only the posts with gaps will participate in propagation of the electromagnetic energy. The other posts, will work as screening medium. 
Thus, by controlling existence of these gaps, one can manipulate with photonic paths on a 2D lattice. Ideally, existence of an oscillator in a certain location of the grid could be controlled {\it in situ} mechanically. This concept of Programmable Cavity Arrays has certain similarities with the well known
Field-Programmable Gate Arrays (FPGA) that revolutionised digital computing and emerging Field-Programmable Analog Array (FPAA).

Fig.~\ref{MPCA} shows electric field distributions for various transmission modes for three examples of the post arrangements. In all cases, the same 8 by 15 array of posts is analysed with full FEM simulation. The grid may be equipped with microwave field source ports on the left and sink ports on the right. In the fist case, a symmetric interferometric system demonstrate both constructive (top) and (destructive) interference. The second array describes an asymmetric two channel system with frequency selective paths. And the final array shows a two input system with two types of interaction. 

The programmable cavity array may be understood as a system performing a linear (empty system) or nonlinear (in presence of nonlinear components inside) transform from $N$ inputs $X_{N}$ to $N$ outputs $Y_{N}$ parametrically dependent on user defined $N$ by $M$ matrix $A_{N,M}$ of bits ($A\in \{0,1\}$): $Y = F(A, X)$, where $N$ and $M$ are a number of raws and columns in the array. By choosing an appropriate matrix $A_{N,M}$ one can design a microwave system for a particular experiment. This approach helps to generalise the process of the microwave system design giving the flexible approach to system building.

\section*{Conclusion}

Multipost re-entrant cavities provide a number of interesting features to the microwave system design. Due to their unique combination of lumped and 3D features of these systems, they can be used in various applications requiring fast and wide range tunability together with the ease of design and prototyping. Combining these important features the multipost re-entrant system provide a new platform for microwave system design.

\section*{Acknowledgments}
This work was supported by the Australian Research Council Grant No. CE110001013.

\hspace{15pt}

\section*{References}

%

\end{document}